\documentclass[]{article}
\usepackage{color}
\usepackage{amsmath,amsfonts,amssymb}
\usepackage{graphicx}
\usepackage{jabbrv}
\usepackage{authblk}
\usepackage{url}
\usepackage{textcomp}
\usepackage{caption}
\usepackage{appendix}
\usepackage{hyperref}
\usepackage{gensymb}
\hypersetup{hidelinks}
\usepackage{geometry}
\title{A hybrid-integrated diode laser in the visible spectral range}

\author[1,$\dag$]{C.A.A. Franken}
\author[1,$\dag$,*]{A. van Rees}
\author[1]{L.V. Winkler}
\author[1]{Y. Fan}
\author[2]{D. Geskus}
\author[2]{R. Dekker}
\author[2]{D.H. Geuzebroek}
\author[3,1]{C. Fallnich}
\author[1]{P.J.M. van der Slot}
\author[1,3]{K.-J. Boller}

\affil[1]{Laser Physics and Nonlinear Optics, Faculty of Science \& Technology, MESA+ Institute, University of Twente, P.O. Box 217, 7500 AE Enschede, the Netherlands}
\affil[2]{LioniX International B.V., P.O. Box 456, 7500 AL Enschede, the Netherlands}
\affil[3]{Optical Technologies Group, Institute of Applied Physics, University of Münster, Corrensstraße 2, 48149 Münster, Germany}
\affil[$\dag$]{Both authors contributed equally to this work.}
\affil[*]{Corresponding author: a.vanrees@utwente.nl}

\begin{document}

\maketitle

\begin{abstract}
Generating visible light with wide tunability and high coherence based on photonic integrated circuits is of high interest for applications in biophotonics, precision metrology and quantum technology.
Here we present the first demonstration of a hybrid-integrated diode laser in the visible spectral range.
Using an \mbox{AlGaInP} optical amplifier coupled to a low-loss Si$\mathbf{_{3}}$N$\mathbf{_{4}}$ feedback circuit based on microring resonators, we obtain a spectral coverage of 10.8~nm around 684.4~nm wavelength with up to 4.8~mW output power.
The measured intrinsic linewidth is $\mathbf{2.3\pm0.2}$~kHz.
\end{abstract}

\section{Introduction}
Integration of diode amplifiers in low-loss passive photonic platforms enables highly coherent light sources in the infrared range~\cite{Fan_2020oe}, where the integrated chip-sized format provides superior stability, portability and scalability for optical systems.
Extending photonic integration into the visible has begun~\cite{Munoz_2019jstqe, Porcel_2019olt} due to its high potential specifically for applications in biophotonics, metrology and quantum technology.
Various optical systems for operation with visible coherent light have already been integrated using the Si$\mathbf{_{3}}$N$\mathbf{_{4}}$ platform~\cite{Munoz_2019jstqe, Porcel_2019olt}, such as for microscopy~\cite{Tinguely_2017oe}, neurophotonic probing~\cite{Sacher_2019oe}, fluorescence biosensing~\cite{Liu_2018bob} or trapping of ions~\cite{Mehta_2020nat}.
Nevertheless, all photonic integrated systems in the visible have so far required external laser sources for their operation, which is due to the lack of hybrid-integrated lasers in this spectral range.

Integrating visible, widely tunable and narrowband lasers would provide several important benefits.
Having the laser integrated on chip removes the instability associated with coupling light into photonic circuits.
Using low-loss and spectrally-selective feedback circuits as integrated extended cavity enhances the laser's spectral stability by narrowing the Schawlow-Townes linewidth on chip~\cite{schawlow_1958pr, boller_2019Phot}.
These advantages carry over to, \textit{e.g.}, refractive index biosensors, which require stable lasers in the visible~\cite{Zinoviev_2011jlt}.
Biosensors based on optical resonance techniques particularly benefit from a narrow linewidth source, because this enhances the spectral resolution of the sensor, and, hence, lowers the detection limit~\cite{White_2008oe}.
Similarly, on-chip operation of multiple-wavelength, highly-coherent visible lasers is important for all-integrated optical cooling and trapping~\cite{Qiao_2019apb} for portable optical clocks~\cite{Poli_2014apb} or for quantum information processing~\cite{Mehta_2020nat}.

In the infrared range, integration of semiconductor amplifiers with spectrally-selective feedback circuits has made various kinds of widely tunable lasers available that can be seamlessly integrated into photonic integrated circuits.
These hybrid and heterogeneously integrated diode lasers have been demonstrated at 2.6~$\mu$m~\cite{Ojanen_2020apl}, 1.65~$\mu$m~\cite{Sia_2020oe}, 1.55~$\mu$m~\cite{boller_2019Phot,Fan_2020oe}, 1.27~$\mu$m~\cite{komljenovic_2015jstqe}, 1.06~$\mu$m~\cite{Bovington_2014oe}, 1.0~$\mu$m~\cite{Zhu_2019oe} and at 0.85~$\mu$m~\cite{Kumari_2017ieeepj}.
For reaching into the blue spectral range, an InGaN-based laser had been epitaxially grown on a Si wafer~\cite{Sun_2016np}, however, the integration into a photonic circuit has not been demonstrated yet.
To avoid two-photon absorption for high photon energies in the visible, photonic circuits require dielectric waveguides with a high band gap, \textit{e.g.}, based on TiO$_{2}$ or Si$_{3}$N$_{4}$.
In addition, suitable waveguides with reduced scattering loss are required, as scattering due to roughness of the core-to-cladding interface strongly increases towards shorter wavelengths ($\sim 1 / \lambda^3$)~\cite{Melati_2014jo}.
Heterogeneous integration, using vertical tapers for evanescent coupling between an InGaN laser and passive TiO$_{2}$ waveguides, was investigated, but successful bonding would require reduced surface roughness~\cite{Kamei_2019pssa}.
To date, no successful realization of a hybrid or heterogeneously integrated diode laser for the visible spectral range has been reported.

In this work we demonstrate the first operation of a hybrid-integrated diode laser in the visible.
Using Si$_{3}$N$_{4}$ waveguides, we provide a feedback circuit with negligible absorption loss and low scattering loss via a weakly confined optical mode.
This hybrid integration enables on-chip narrow-linewidth lasers in the visible spectral range.

\section{Laser design and hybrid integration}

\begin{figure}[htb]
	\centering
	\includegraphics[width=0.7\linewidth]{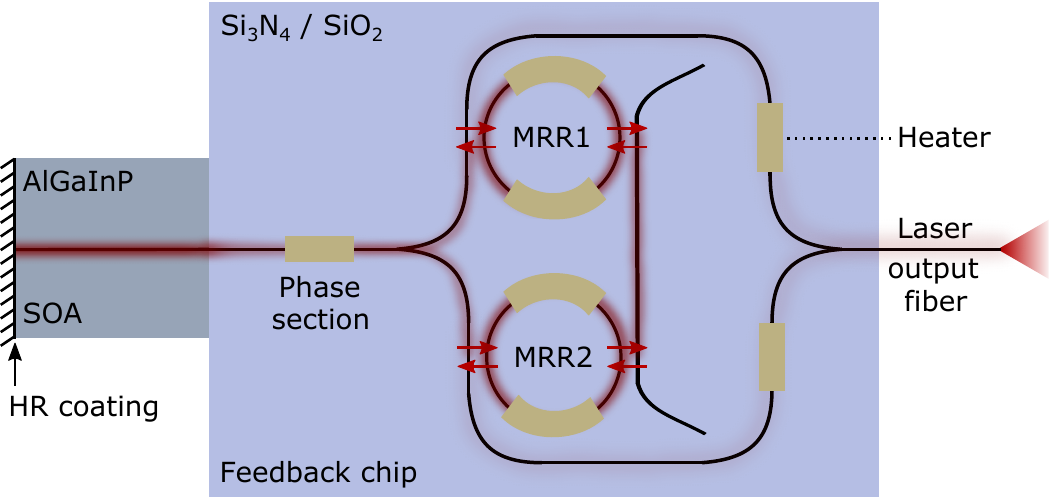}
	\caption{Schematic overview of the hybrid laser comprising an \mbox{AlGaInP} optical amplifier butt-coupled to a Si$_3$N$_4$/SiO$_2$ feedback chip. The chip contains two sequential microring resonators (MRR1, MRR2) that form a Vernier filter within a loop mirror. Directional couplers are indicated by red arrows. Heaters (yellow) are placed on the MRRs, on the intracavity phase section and near the output Y-junction.}
	\label{fig:setup}
\end{figure}

The design of the hybrid laser is shown schematically in Fig.~\ref{fig:setup}.
The laser cavity is formed by a semiconductor chip coupled to a Si$_{3}$N$_{4}$/SiO$_{2}$ based feedback chip.
The multiple quantum well AlGaInP semiconductor optical amplifier (SOA, Sacher SAL-0690-25) is 600~$\mu$m long and provides a gain bandwidth of approximately 10~nm around 685~nm, with up to 25~mW output power in Littrow configuration.
Its back facet has a high reflectivity (HR) coating of 95\%, which forms one mirror of the laser cavity.
The other mirror is a reflective Vernier filter formed by a y-junction and two cascaded microring resonators (MRRs) to provide highly frequency-selective feedback.

To minimize propagation losses and to support only single-mode propagation, the feedback circuit is implemented using a high aspect ratio Si$_{3}$N$_{4}$ core that is 2~$\mu$m wide and 25~nm thick.
This core is centered in a 16~$\mu$m thick SiO$_{2}$ cladding to guide the TE$_{00}$ mode~\cite{roeloffzen_2018jstqe,Porcel_2019olt}.
The smallest bending radius for this waveguide geometry is chosen as 1200~$\mu$m to ensure that bend radiation loss is small compared to the straight propagation loss.
Key to low propagation loss is the small height of the sidewalls to minimize the scattering loss, as the sidewall etching produces a larger interface roughness than the layer deposition process, which defines the top and bottom surface of the waveguide.
To characterize the straight-waveguide propagation loss, we measured the light transmission through 1-cm and 4-cm long straight waveguides and 58-cm long low-curvature spirals.
The measured propagation loss is 0.07$\pm$0.02 dB/cm at the wavelength of 685~nm.
To obtain the overall propagation loss of the rings, we scanned a tunable laser across several ring resonances.
This resonator loss is 0.13$\pm$0.02 dB/cm, based on fitting the calculated ring responses to the measured transmission.
The excess resonator loss is probably due to bending loss and parasitic loss in the symmetric coupler section~\cite{spencer_2014o}.
In order to optimize the mode matching with a single-mode, polarization-maintaining output fiber (Nufern PM460-HP), the waveguide width is tapered down to 0.8~$\mu$m at the fiber side to obtain a simulated coupling efficiency of 85\%.
The waveguide is not tapered at the gain side, as calculations predicted a comparably high overlap of 89\% with the estimated mode of the SOA (MFD$_{hor}$ = 3.6~$\mu$m, MFD$_{vert}$ = 1.5~$\mu$m).

The functionality of the feedback circuit is twofold.
Frequency filtering is realized using two high-quality MRRs with slightly different radii.
These rings have radii of 1200 and 1205~$\mu$m and an estimated loaded Q-factor of $9.5 \cdot 10^5$.
The measured Vernier free spectral range (FSR) is 9.90$\pm$0.05~nm at the nominal wavelength of 685~nm~\cite{boller_2019Phot}.
This FSR is approximately equal to the gain bandwidth of the SOA, to avoid multiple wavelength oscillation.
The second function is to lower the intrinsic laser linewidth with a long effective length of the feedback circuit~\cite{liu_2001apl}.
At resonance, the effective length of each ring is enhanced by a factor of 18 to 14~cm, which is calculated by the group delay, using the ring radius, the measured resonator loss and the power coupling between bus waveguide and ring.
This power coupling $\kappa^2=0.043\pm0.011$ is measured using a separate directional coupler, identical to the ones in the rings.

Summing the optical lengths of all elements gives a maximum effective roundtrip length of 49~cm for the laser cavity.
As the length of the laser cavity is strongly frequency-dependent, the cavity modes are non-equidistant with a calculated minimum mode spacing of 1.0~pm (0.67~GHz).
Single longitudinal mode oscillation is expected, because this mode spacing is larger than the calculated 0.46~pm (0.30~GHz) full-width-at-half-maximum (FWHM) of the Vernier resonance.

Thermo-optic phase tuning of the feedback circuit is realised via chromium-based~\cite{roeloffzen_2018jstqe} resistive heaters.
Tuning the ring heaters is used for wavelength selection, while tuning the intracavity phase section is used to align a cavity resonance with the ring resonances.
The light extracted from the cavity is combined using a Y-junction before coupling it into the output fiber. 
To balance the phase in both extraction paths with minimal thermal tuning for maximum output coupling, one of both heaters placed near this Y-junction can be tuned.
Introducing a phase shift of 2$\pi$ with the resistive heaters requires approximately 400~mW of dissipated electrical power.

To enable stable laser operation via hybrid integration, the amplifier, feedback chip and output fiber were all aligned for optimum coupling and bonded with an adhesive.
The SOA and heaters were wire bonded to an adapter board for electrical connections.
The SOA pump current and heater currents were supplied with a Thorlabs LDC205B current source and a Nicslab XPOW-8AX-CCvCV multi-channel power controller, respectively. 
To remove the generated heat, the hybrid assembly was placed on a common subcarrier with a heat sink, which was temperature-controlled using an adequately dimensioned Peltier element.
For all measurements, the temperature was set at 20~\degree~C.

\section{Experimental results}

\begin{figure}[htb]
\centering
	\includegraphics[width=0.7\linewidth]{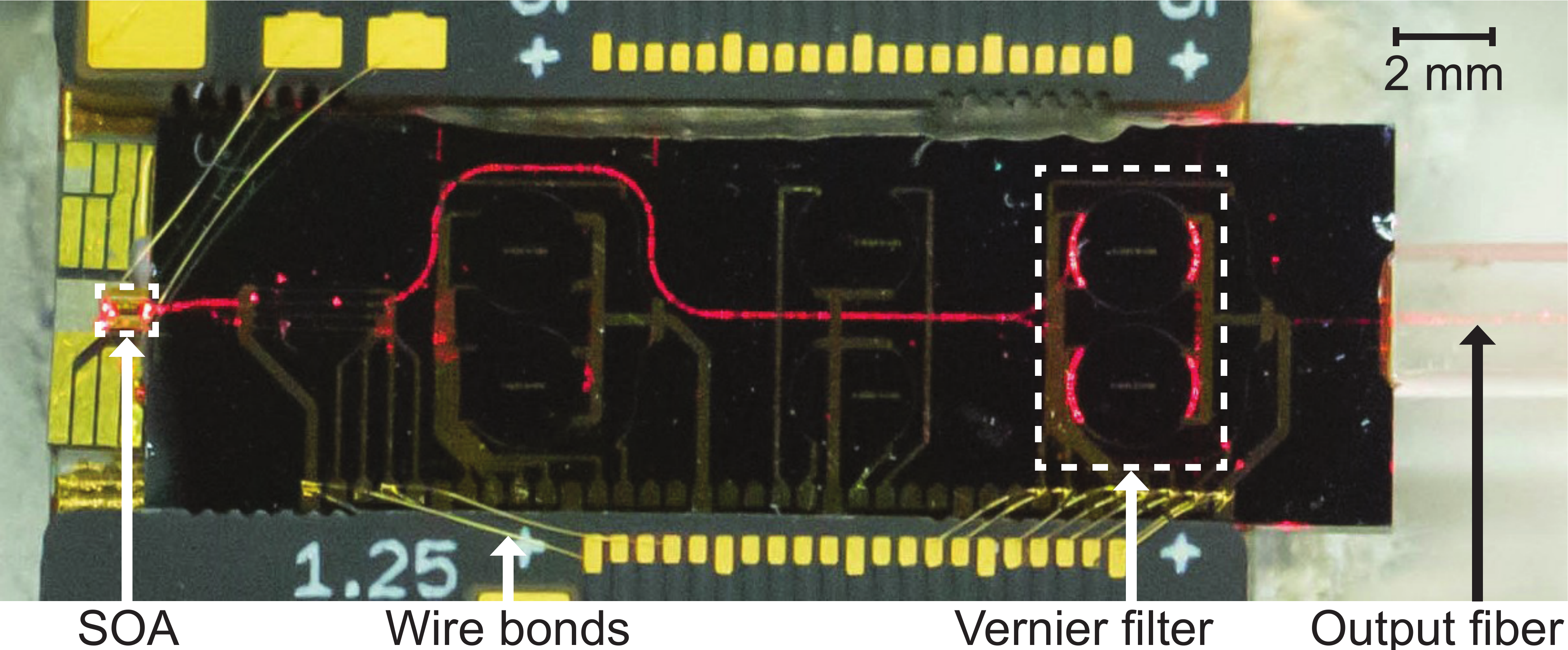}
	\caption{
	Photograph of the hybrid laser pumped at 90 mA. Bright scattered light from both rings indicates that a cavity mode is tuned to be aligned with both ring resonances. This light is partly blocked by heaters on top of the rings.}
	\label{fig:hybridassembly}
\end{figure}

When pumping the SOA above treshold, scattered red light from the feedback chip can be observed.
Figure~\ref{fig:hybridassembly} shows a photograph of the assembled laser when pumped at the maximum allowed pump current of 90~mA.
This photo demonstrates the first operation of a hybrid-integrated diode laser in the visible spectral range.
We tuned the phase section to align a cavity resonance with a Vernier resonance of the two rings, which is confirmed by the bright scattered light from both rings.

\begin{figure}[htb]
	\centering
	\includegraphics[width=0.35\linewidth]{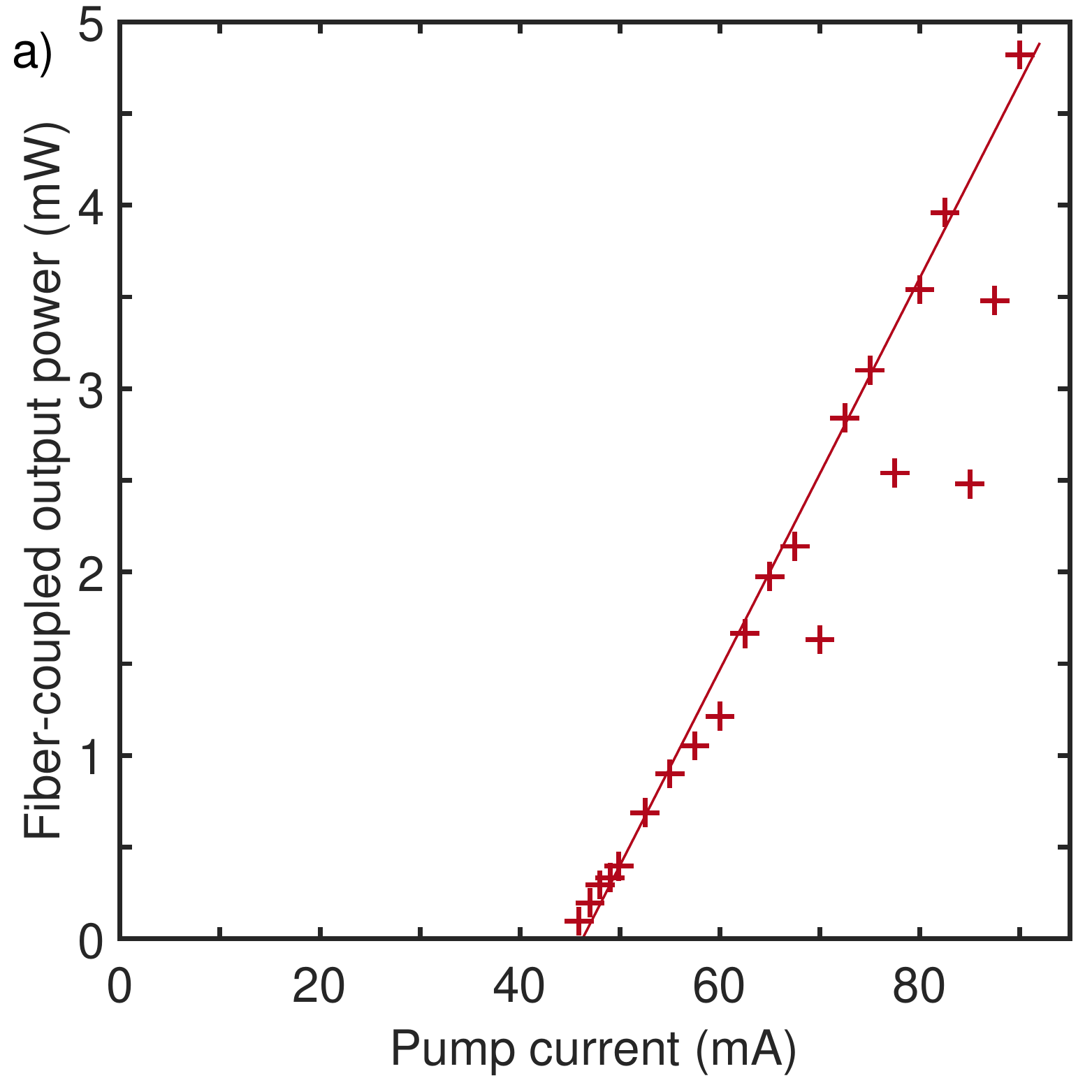}
	\includegraphics[width=0.35\linewidth]{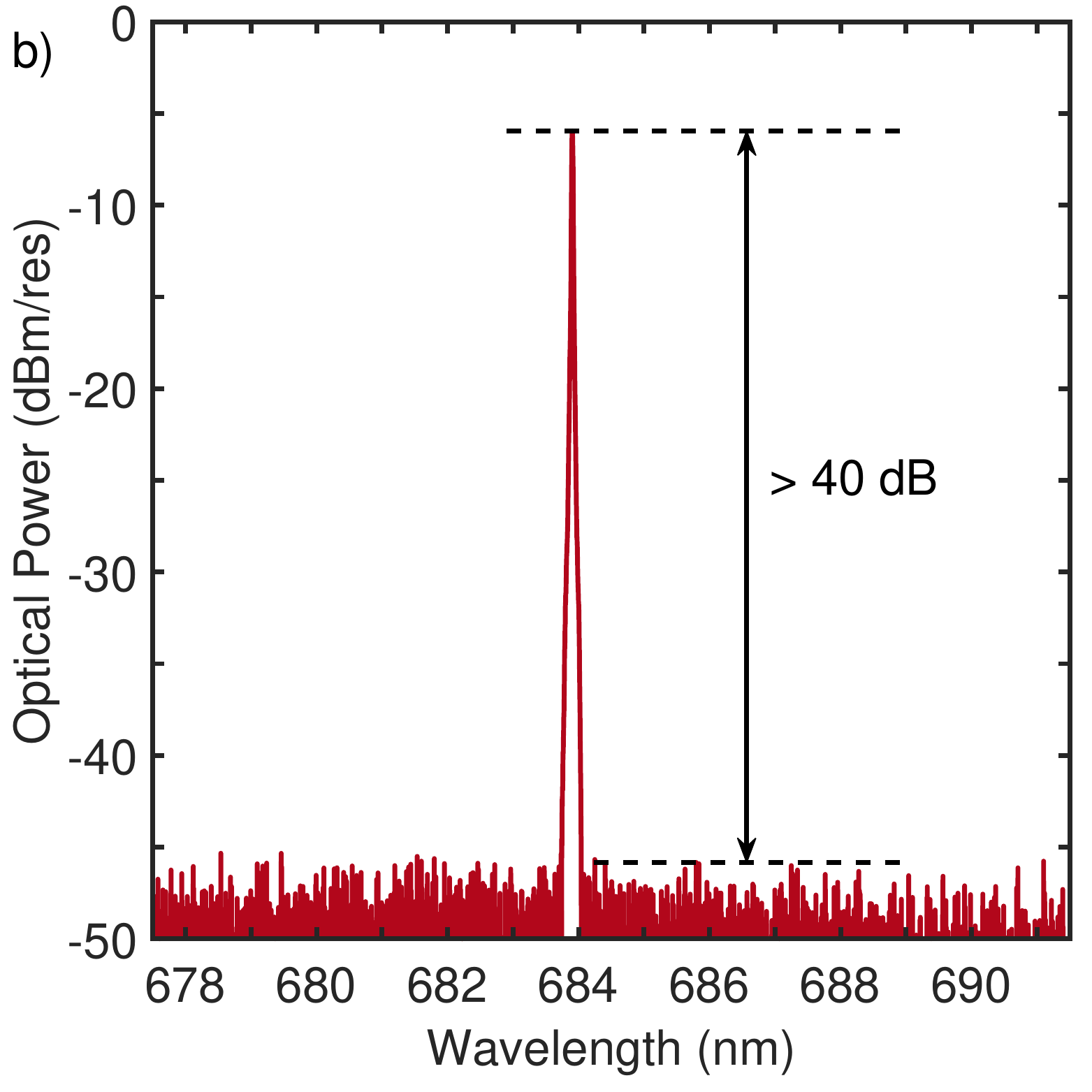}
	\caption{
	a) Fiber-coupled output power versus pump current, with optimization of the phase section heater.
	b) Optical power spectrum measured for a pump current of 90~mA.}
	\label{fig:power}
\end{figure}

To measure the laser output power and spectral characteristics, the laser was connected via a fiber coupler (Thorlabs TW670R5A2) to a power sensor and an optical spectrum analyzer (OSA, ANDO AQ6317). 
Figure~\ref{fig:power}(a) shows the measured fiber-coupled output power as function of pump current, corrected for the 50:50 splitting ratio of the fiber coupler.
For each measurement, the heater on the phase section was optimized for maximum output power, while other heaters were not activated.
Figure~\ref{fig:power}(a) indicates a threshold current of 46~mA.
Above threshold, the output power increases approximately linearly with a slope efficiency of 0.11 mW/mA, indicated with the linear fit line through all points except the four low outliers.
These outliers were likely caused by spectral mode hops, since current-induced refractive index changes of the SOA caused the laser to hop to modes that were less efficiently coupled out.
We measured a maximum output power of 4.8~mW at the maximum specified current of 90~mA.
For this setting, we also measured the laser's output spectrum with the OSA using a 0.01~nm resolution bandwidth.
This spectrum, as shown in Fig.~\ref{fig:power}(b), reveals a resolution-limited single-wavelength peak at 683.9~nm with a high signal-to-noise ratio of 40~dB.
As the resolution bandwidth of the OSA is insufficient to distinguish adjacent cavity modes, additional measurements are described below to confirm single-mode oscillation.

\begin{figure}[htb]
	\centering
	\includegraphics[width=0.35\linewidth]{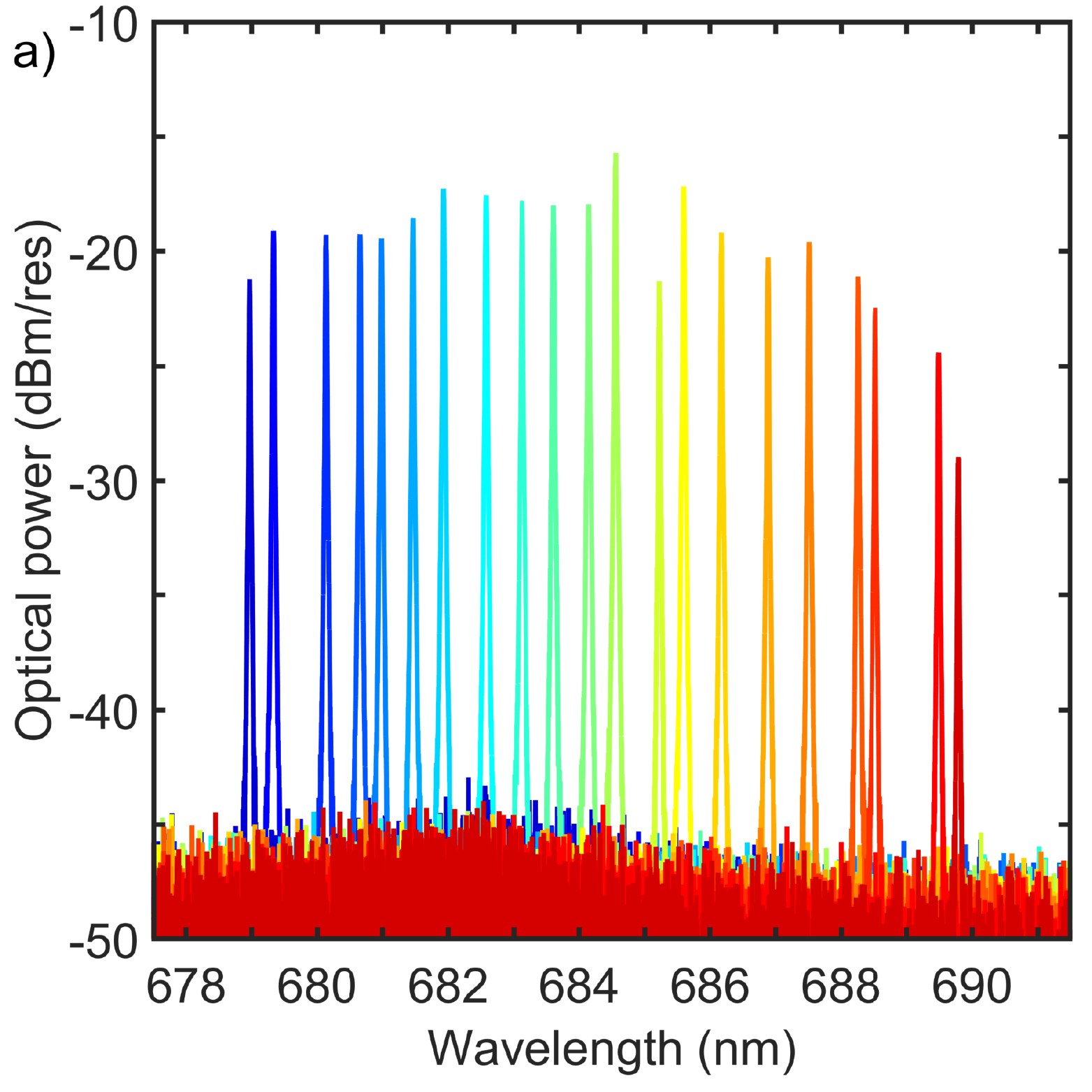}
	\includegraphics[width=0.35\linewidth]{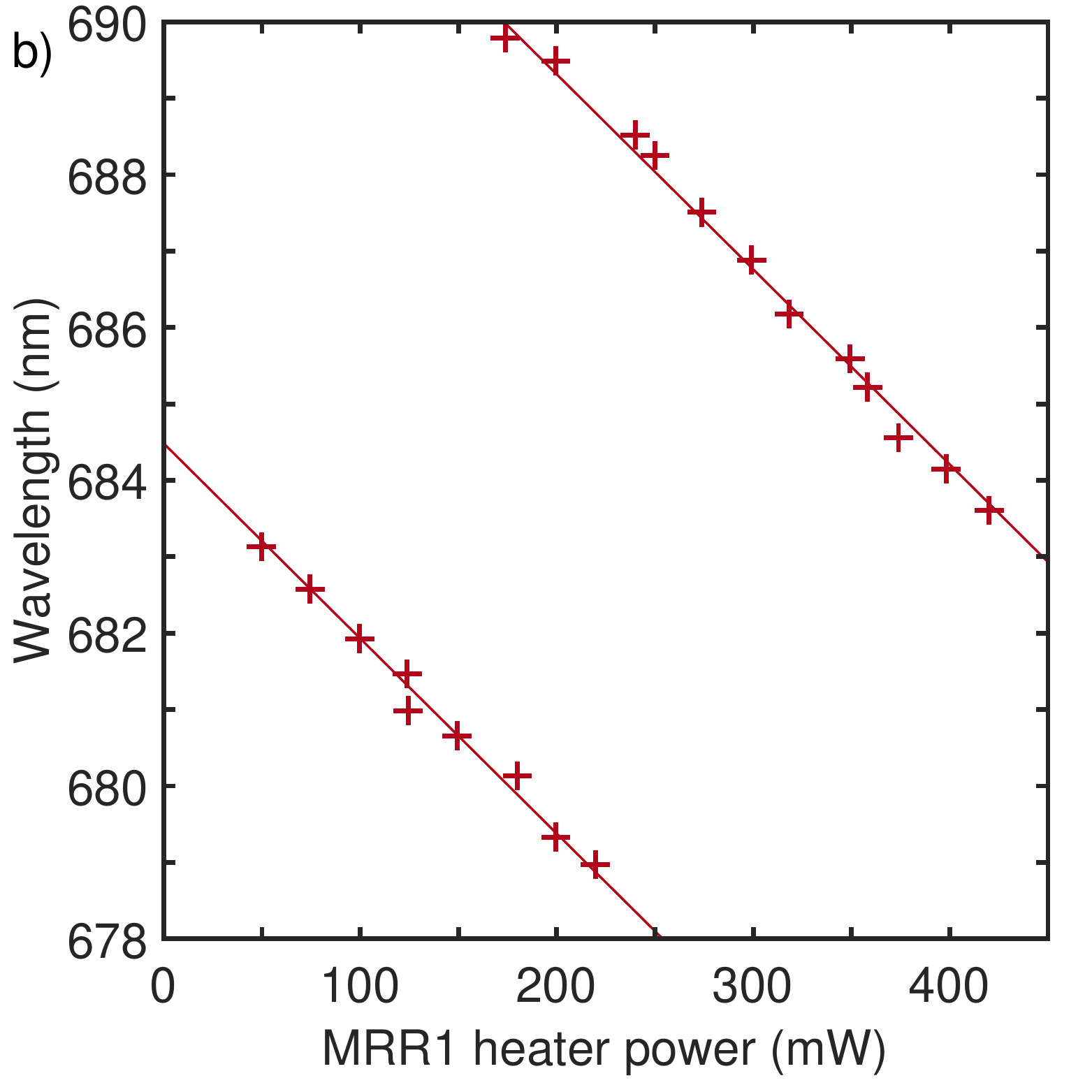}
	\caption{
	a) Superimposed laser spectra, as measured with an OSA set to 0.01~nm resolution and obtained by varying the heater power to MRR1.
	b) Peak laser wavelength extracted from the laser spectra in a) as function of heater power.}
	\label{fig:tuning}
\end{figure}

To determine the spectral coverage of the laser, we varied the electrical power to the heater on MRR1.
This shifts the resonant frequency of the underlying ring, thereby changing the Vernier feedback frequency.
Fig.~\ref{fig:tuning}(a) shows several superimposed laser spectra as measured with the OSA.
The pump current was set to 50~mA to record the spectral coverage already obtainable with a few mA above threshold.
The phase section and a heater near the output Y-junction were coarsely optimized for each measurement to obtain single-wavelength laser emission with optimum output power.
We observe that the laser wavelength can be tuned, in discrete steps, over a range of 10.8~nm around 684.4~nm, which covers the entire gain bandwidth of the amplifier.
The highest output power was measured near the peak of the gain spectrum, as expected.
Further variation in power can be explained by the nonuniform gain spectrum and by nonoptimal heater settings, since the outcoupling ratio strongly depends on the fine-tuning of the ring resonances. 
For a better qualification of laser tuning, Fig.~\ref{fig:tuning}(b) shows the extracted peak wavelengths as function of the heater power at MRR1.
The laser wavelength decreases approximately linearly with increasing heater power, except for large hops when the laser is operated near the edge of the gain bandwidth.
We fitted two parallel lines through the data points to retrieve the hop distance.
The wavelength spacing between these lines is 9.9~nm, which agrees well with the separately measured FSR of the solitary Vernier filter.

\begin{figure}[htbp]
	\centering
	\includegraphics[width=0.35\linewidth]{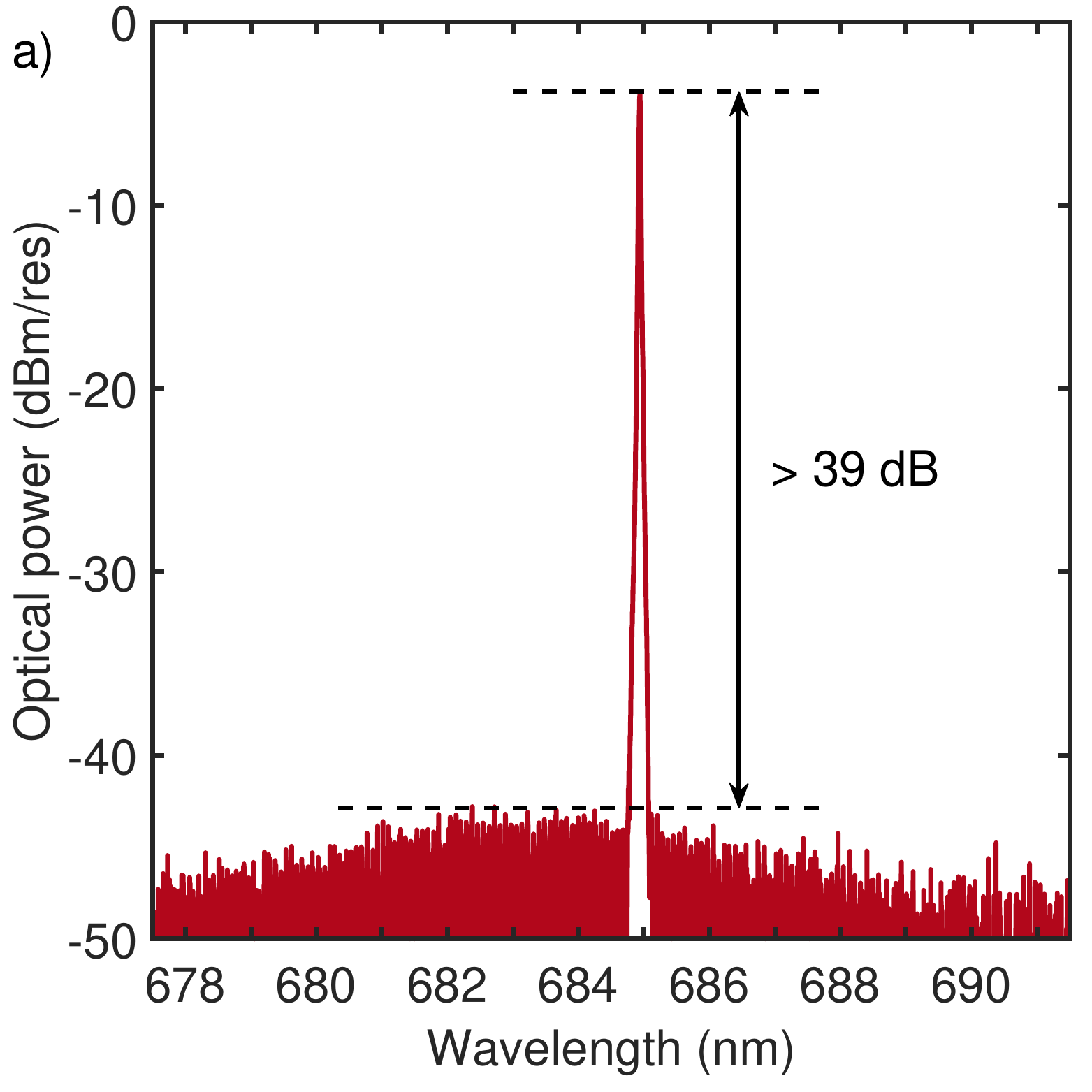}
	\includegraphics[width=0.35\linewidth]{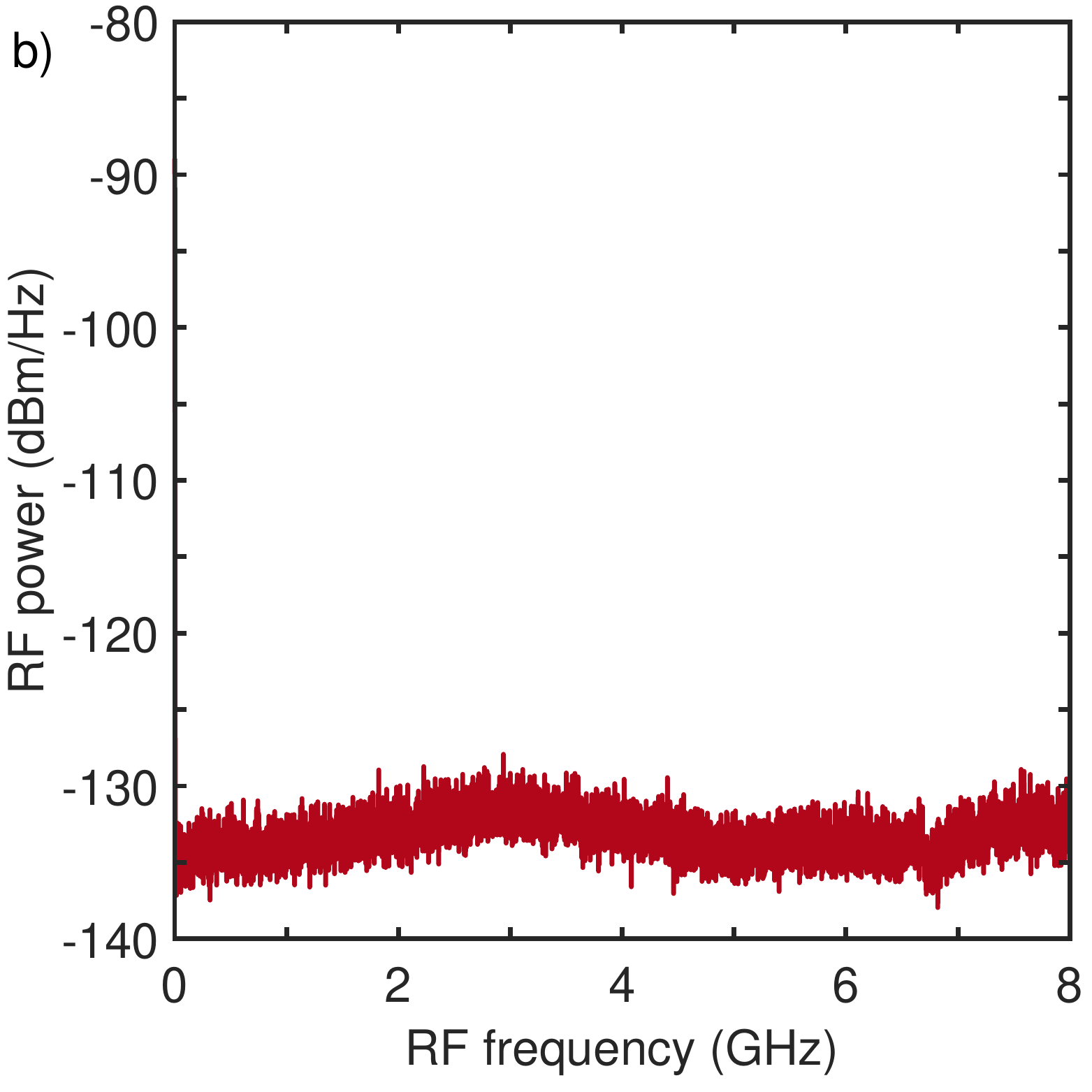}
	\caption{
	Optical spectrum (a) and RF spectrum (b) of the laser output with the laser optimized for single longitudinal mode oscillation.
	The pump current was set to 90~mA.}
	\label{fig:Beats}
\end{figure}

To verify that optimized heater settings provide oscillation in a single longitudinal mode, \textit{i.e.}, to spectrally resolve also adjacent longitudinal modes, we used a setup to detect the beating of laser modes in the radio frequency (RF) domain.
For this purpose, the laser output was sent through an isolator (Thorlabs IO-F-690APC) directly to the OSA and directly to a fast photodiode (Thorlabs DXM12CF) connected to an RF spectrum analyzer (RFSA, Agilent E4405B).
The RFSA was set to a resolution bandwidth of 1~MHz, and swept over a range up to 8~GHz, which would be sufficient to detect the beating between multiple longitudinal modes within a single Vernier resonance.
We emphasize that the RF range is wider than the 0.01~nm (6~GHz) resolution bandwidth of the OSA, which ensures that multiple mode oscillations would be noticed.
Figure~\ref{fig:Beats} shows an example of the recorded (a) optical and (b) RF spectra, when the laser was optimized for single longitudinal mode oscillation.
This optical spectrum shows a single wavelength, while corresponding RF spectra, recorded before and after this OSA measurement, show no indication of the presence of any beat frequencies.
We note that several beat frequencies showed up in the RF spectrum with non-optimized settings of the laser.
The RF and optical spectra combined let us conclude that, with proper settings, the laser oscillates in a single longitudinal mode with a side mode suppression ratio of at least 39~dB.

\begin{figure}[htbp]
	\centering
	\includegraphics[width=0.525\linewidth]{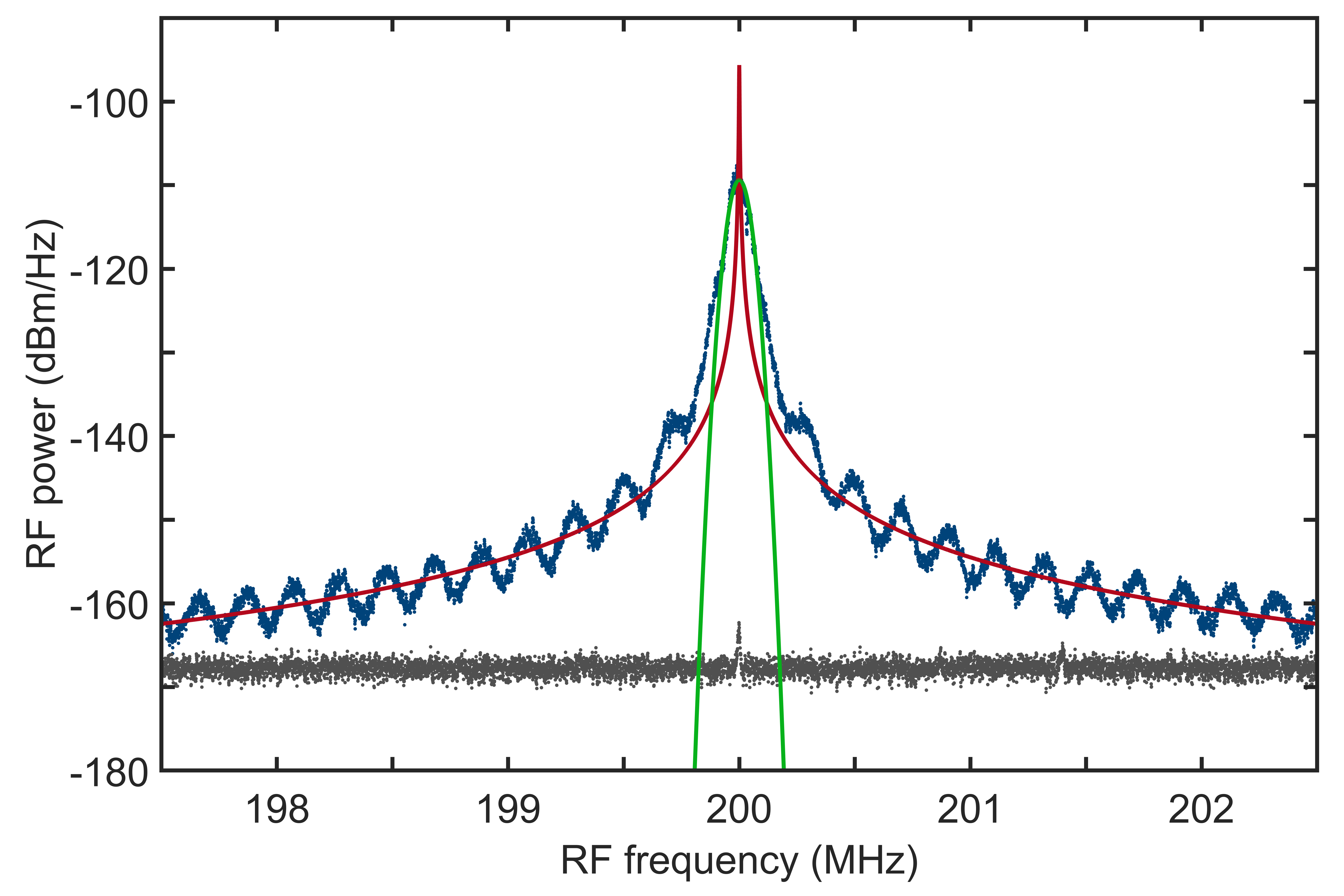}
	\caption{
	Recorded beat signal (blue trace) and background (grey trace) from the delayed self-heterodyne measurement setup, together with a fitted Gaussian profile (green curve) and a fitted Lorentz profile (red curve).}
	\label{fig:Linewidth}
\end{figure}

To determine the intrinsic linewidth of the laser, the output was sent through an isolator to a delayed self-heterodyne detection setup.
The setup uses a fiber-based Mach-Zehnder interferometer, where one arm contains a 200~MHz acousto-optic modulator, while the other arm contains a 1-km long single-mode fiber as a delay (Thorlabs SM600), limited in length because of 6~dB/km propagation loss. 
The beat signal was recorded using the fast photodiode connected to the RFSA, set to a radio and video bandwidth of 10~kHz, and averaged over 50 sweeps.
Fig.~\ref{fig:Linewidth} shows the measured power spectrum as the blue trace.
For this measurement, the pump current was supplied by a battery-operated current source (ILX Lightwave LDX-3620) set to 90~mA.
We tuned MRR2 to set the laser's wavelength near maximum gain at 685.6~nm and fine-tuned the phase section to minimize the intrinsic linewidth.
The recorded RF signal is well above the noise floor of the setup, displayed as the grey trace.
In the blue trace, periodic modulations of approximately 200~kHz can be observed, consistent with the 1-km fiber delay and an even longer coherence length of the laser~\cite{richter_1986jqe}.
By fitting this trace to a Gaussian profile, as shown with the green curve, we find that the central peak has a predominantly Gaussian line shape with a FWHM of 56.3$\pm$0.4~kHz.
This line shape of the free-running laser, which is associated with $1/f$ frequency noise~\cite{mercer_1991jlt} caused by technical noise sources, can be narrowed by locking the laser frequency to a high-finesse cavity or a Sr transition~\cite{Qiao_2019apb}.
To estimate the intrinsic linewidth, which arises from white frequency noise, we fit the wings of the recorded signal to a Lorentz function, as shown with the red curve.
The fitted Lorentzian linewidth is 2.3$\pm$0.2~kHz, which confirms that the laser has a very narrow intrinsic linewidth.

\section{Conclusion}

To conclude, we demonstrate the realization of the first hybrid-integrated diode laser in the visible spectral range.
The maximum output power is 4.8~mW for a pump current of 90~mA and the spectral coverage amounts to 10.8 nm around 684.4~nm.
The hybrid integration of an \mbox{AlGaInP} optical amplifier with a Si$_{3}$N$_{4}$ feedback chip, used for sharp Vernier filtering with high-quality microring resonators, enables single longitudinal mode laser oscillation.
The long effective optical cavity roundtrip length of up to 49 cm provides a narrow intrinsic linewidth of down to 2.3$\pm$0.2~kHz.
Improving the laser design, \textit{e.g.}, by increasing the core thickness, would enable a smaller bend radius and a thinner top cladding for more efficient heaters. Second, adding an intracavity tunable coupler for the extraction of the output power independent from the frequency filtering~\cite{boller_2019Phot}, would improve the overall performance of the laser, including more spectrally uniform and higher output power and lower intrinsic linewidth.
The concept of hybrid-integrated diode lasers with very small intrinsic linewidth could possibly cover the whole visible spectral range.
Realization of a fully integrated multi-color visible laser engine~\cite{Mashayekh_2021oe} would then become feasible for numerous applications.

\section*{Funding}
European Union’s Horizon 2020 research and innovation program (688519) (PIX4life).

\section*{Acknowledgments}
The authors would like to thank R. Heuvink for support with the feedback circuit characterization.

\section*{Disclosures}
The authors declare no conflict of interest.

\bibliographystyle{osajnl}

\end{document}